\documentclass{nature}
\usepackage{amsmath,amssymb}    
\usepackage{graphicx}           
\usepackage{color}

\title{Multiple peaks patterns of epidemic spreading in multi-layer networks}
\author{Muhua Zheng$^{1}$, Wei Wang$^{2}$, Ming Tang$^{3}$, Jie Zhou$^{1}$, S. Boccaletti$^{4,5}$ and Zonghua Liu$^1$$^*$}

\begin{document}

\maketitle

\begin{affiliations}
\item Department of Physics, East China Normal University,
Shanghai, 200241, China

(* zhliu@phy.ecnu.edu.cn)

\item Web Sciences Center, University of Electronic
Science and Technology of China, Chengdu 610054, China

\item School of Information Science and Technology, East China Normal University,
Shanghai, 200241, China

\item CNR-Institute of Complex Systems, Via Madonna del Piano, 10, 50019 Sesto Fiorentino, Florence, Italy

\item The Embassy of Italy in Tel Aviv, 25 Hamered Street, 68125 Tel Aviv, Israel

\end{affiliations}

\begin{abstract}
The study of epidemic spreading on populations of networked individuals has seen recently a great deal of
significant progresses. A common point of all past studies is, however, that there is only one peak
of infected density in each single epidemic spreading episode. At variance, real data from different cities over
the world suggest that, besides a major single peak trait of infected density, a finite probability exists
for a pattern made of two (or multiple) peaks. We show that such a latter feature is fully distinctive
of a multilayered network of interactions, and reveal that actually a two peaks pattern emerges from different time
delays at which the epidemic spreads in between the two layers. Further, we show that essential ingredients are
different degree distributions in the two layers and a weak coupling condition between the layers themselves.
Moreover, an edge-based theory is developed which fully explains all numerical results. Our findings may therefore
be of significance for protecting secondary disasters of epidemics, which are definitely undesired in real life.
\end{abstract}

Epidemic spreading in networked populations has been studied intensely in the last decade, and a lot of great
progresses has been achieved \cite{Pastor:2015,Barrat:2008,Wang:2017,Salehi:2015} which significantly
increased our understanding. This is actually useful for public health authorities to assess
situations quickly, to take and enforce informed decisions, and to optimize vaccination and drug delivery policies.
Initially, the main
attention focused on static networks, where each node represents an immobile agent and the contagion
occurs only between neighboring nodes: it was remarkably revealed that scale-free
networks display a vanishingly small epidemic threshold in the thermodynamic limit
\cite{Pastor-Satorras:2001,Boguna:2002}. After that, the focus shifted to reaction-diffusion models \cite{Colizza:2007a,Colizza:2007b,Andrea:2008,Liu:2009}, flow-driven epidemics
\cite{Vazquez:2007,Meloni:2009,Balcan:2009,Ruan:2012,Liu:2012}, objective spreading \cite{Tang:2009,Liu:2010}
and adaptive behaviors \cite{Gross:2006,Gross:2008,Zhou:2012,Zhou:2013,Marceau:2010}. Finally, in a third stage,
multilayered
\cite{Boccaletti:2014,Feng:2015,Sahneh:2013,Wang:2013,Yagan:2013,Newman:2005,Marceau:2011,Zhao:2014,
Buono:2015,Buono:2014} and temporal \cite{Holme:2012,Perra:2012} networks were assumed to play a critical role
on such processes.

A common feature in past studies is the use of phenomenological models which produce
a single peak of infected density in each individual epidemic spreading. An interesting
question is therefore whether or not all real evolutionary processes are conveniently represented by such
a framework.
A scrupulous analysis of a large number of real data from different cities over
the world surprisingly shows that, besides a major pattern made of a single peak,
there is a finite and non negligible probability for a new pattern of epidemic outbreak
featuring two (or multiple) peaks. Notice that a two peaks pattern
implies two outbreaks in a single spreading period, i.e. a secondary
occurrence of the same epidemics, which may in turn produce severe calamities
and disasters within unprepared populations. Understanding
the underlying mechanism at the basis of this new pattern (with the help of a novel model
extracted from real data) is therefore quintessential to properly cope with such life-threatening hazards.

We proceed by making three steps: the first is to
build a suitable model for the data, the second is to reproduce the data by the
 model, and the third is to suggest effective ways for predictions and control of the
epidemic spreading.  As social activities and interactions occur in
network structures, we here consider the epidemics in different geographic regions (or cities) as that occurring in
multilayered graphs. Namely, we will take two coupled neighboring regions as an example, and construct a
two-layered network model which fully reproduces the observed patterns of epidemics. In particular, we
demonstrate that the pattern of two peaks originates from large time delays of epidemic outbreaks between
the two layers, which depends in turn on both the difference in the degree distributions of the two layers
and a weak coupling condition between them. To better understand the findings, an edge-based theory is developed
which perfectly agree with the numerical simulations.

\section*{Results}

\subsection{Patterns of epidemic outbreak with two or multiple peaks in real data.}
Monitoring the potential outbreaks of an epidemic spreading is of extreme importance for protection of our society.
Based on the detected trend of spreading of infections such as SARS (Severe Acute Respiratory Syndrome), H1N1 (Swine Influenza), H5H1 (Avian Influenza), and Ebola, one can indeed attempt to enforce suitable measures able to reduce the epidemic at
its maximum extent.  For this purpose, many countries have established their sentinel surveillance systems to collect epidemic
data. For instance, Hong Kong Department of Health has organized a surveillance system, with the aim of collecting empirical
data of infectious diseases, and of analyzing and predicting the trend of the infection. In such a system,
there are about $64$ General Out-Patient Clinics
(GOPC) and $50$ General Practitioners (GP), which form two distinct sentinel surveillance networks of the city
\cite{www:hongkong,Zheng:2015}. In these two networks, one obtains for instance the weekly consultation rates
of influenza-like illness (per $1,000$ consultations), which reflect the overall influenza-like illness activity
in Hong Kong.

Fig. 1(a) shows the collected data from $1998/1/3$ to $2014/8/2$ in GP, while the corresponding
data of GOPC is shown in Fig. 1 of the Supplementary Information (SI). From Fig. 1(a) one see that there are many events
of epidemic spreading, and the intervals between
two consecutive events are not regular, indicating non-periodic outbreaks of recurrent epidemics
\cite{Zheng:2015}. On the other hand, one notices from Fig. 1(a) that most of the outbreaks correspond to
a single peak of infected density, which is the pattern well described by the classical susceptible-infected-refractory
(SIR) models. However, one notices also that there is a finite
probability for a novel pattern of epidemic outbreak which features, instead, two or multiple peaks. Fig. 1(b) shows one of such
patterns (with two peaks) occurring at around $2005/6$, which indicates that an infectious disease raised two times
during that epidemic period in Hong Kong. Such unexpected phenomenon also exists in the data from GOPC (as one
can see in  Fig. 1 of the SI).

These multiple peaks patterns are actually occurring generically, and are not limited to a specific geographical region.
Remrkably, indeed, epidemic data
from other sources and cities display ubiquitously patterns similar to that reported in Fig. 1(b).
For instance, Fig. 1(c) shows the data of the weekly measles infective cases (WMICs) from 1908 to 1937 in Boston
\cite{measles:2016,Scarpino:2016}, and a two peaks pattern is shown in Fig. 1(d) at around 1915/4. In
addition, two or multiple peaks of infected density may characterize the outbreaks in the total number of WMICs
of two neighboring cities, also when uni-modal patterns are actually observed in each individual city. For example,
Fig. 1(e) and (f) report the data of WMICs in Bristol and Newcastle \cite{www:measdata}:  the yellow line denotes the
total number of WMICs, whereas the blue and green lines represent the data in Bristol and Newcastle,
respectively. In Fig. 1(f) one can well appreciate that the pattern of two peaks occurs only in the total
number of WMICs. Similarly, Fig. 1(g) and (h) show the case of Bristol and Sheffield \cite{www:measdata}, and once again
a typical two peaks pattern [Fig. 1(h)] occurs.

\subsection{The two-layered network model.}
To capture the underlying mechanism, we introduce a model of a
two-layered network, where the two layers represent actually two interconnected regions or cities.
Fig. 2 is a sketch of the model: $\mathcal{A}$ and $\mathcal{B}$ are the two layers,
which are coupled through the inter-network $\mathcal{AB}$. For
the sake of simplicity, we let the two networks $\mathcal{A}$ and $\mathcal{B}$ have the same size $N_a=N_b$.
Furthermore,  $\langle k_a\rangle$, $\langle k_b\rangle$, and $\langle k_{ab}\rangle$ represent the average degrees
of networks $\mathcal{A}$, $\mathcal{B}$ and $\mathcal{AB}$, respectively (see {\sl Methods} for details).

Each node is a unit of the SIR model, where $S, I$ and $R$ represent the susceptible, infected
and refractory phases of individuals, respectively. At each time step, a susceptible node will be infected by an
infected neighbor with rate $\beta$, and an infected node will become refractory with probability $\mu$. The
infectious process will be considered terminated when no more infected nodes exist. While $\mu$ is taken to be the same for all networks,
we let $\beta_a$, $\beta_b$ and $\beta_{ab}$ be the infectious rates of networks $\mathcal{A}$,
$\mathcal{B}$ and $\mathcal{AB}$, respectively.

A pattern of two peaks appears in the numerical simulations of our model. We choose $\mathcal{A}$ to
be a scale-free (SF) network with degree distribution $P_A(k)\sim k^{-\gamma}$ \cite{Catanzaro:2005}, and
$\mathcal{B}$ a random regular (RR) network with a constant degree $k_b$ \cite{Rucinski:2012}. The inter-network
$\mathcal{AB}$ is constructed by randomly adding links between $\mathcal{A}$ and $\mathcal{B}$ until an averaged
degree $\langle k_{ab}\rangle$ is attained. In simulations, we fix $N_a=N_b=10\,000$, and set initially $0.1\%$ of the
individuals in $\mathcal{A}$ to be infected. The yellow circles in Fig. 3(a) report the evolution of the
infected density $\rho_I$ in the whole network with the parameters $\gamma=2.1$,
$\langle k_a\rangle=\langle k_b\rangle=6$, $\langle k_{ab}\rangle=1.0$, $\beta_a=\beta_b=0.05$,
$\beta_{ab}=0.005$ and $\mu=0.1$. One can easily differentiate two peaks of $\rho_I$, indicating that the
empirical observations in Fig. 1 can be fully reproduced.

To gather a deeper understanding,
we also measure the evolutions of $\rho_I^A$ and $\rho_I^B$ in both layers $\mathcal{A}$
and $\mathcal{B}$, and report them as green triangles and  blue squares in Fig. 3(a), respectively. It
is easy to see that the times at which the maximum infected density is obtained in layers $\mathcal{A}$ and
$\mathcal{B}$ are different, implying that the pattern of two peaks is likely triggered by the time difference of
epidemic outbreaks in the two layers.

The next step is to focus on the key factors that determines the occurrence of such two peaks pattern.
For this purpose, we first concentrate on the role of the average degree $\langle k_{ab}\rangle$ of the inter-network
$\mathcal{AB}$. Fig. 3(b) shows the evolution of the infected density $\rho_I$ with different $\langle k_{ab}\rangle$
(with triangles, squares and circles denoting the cases of $\langle k_{ab}\rangle=3.0, 1.0$ and $0.5$,
respectively). One notices that the pattern of $\rho_I$ is uni-modal when $\langle k_{ab}\rangle$ is large, but
bimodal when $\langle k_{ab}\rangle$ is sufficiently small, indicating that $\langle k_{ab}\rangle$ is a key factor
for the appearance of a bimodal pattern: a smaller $\langle k_{ab}\rangle$ favours the appearance of the two peaks
pattern, indicating that the two main networks $\mathcal{A}$ and $\mathcal{B}$ should be only weakly coupled among
them.

As a second step, we study the influence of the infectious rate $\beta_{ab}$ on the pattern.
Fig. 3(c) reports the results obtained for different $\beta_{ab}$ (with triangles, squares and circles
representing the cases of $\beta_{ab}=0.05, 0.005$ and $0.001$, respectively). Once again one may notice that
the condition of a weak coupling is essential for a bimodal pattern: $\rho_I$ is indeed
uni-modal when $\beta_{ab}$ is large, while the two peaks appear when $\beta_{ab}$ is small.
Finally, we study the influence of the exponent $\gamma$ of
the SF network. Fig. 3(d) shows that the bimodal feature is reduced with the increase of
$\gamma$. As a larger $\gamma$ means a smaller difference between the structures of the SF and RR
networks, one can infer that also the heterogeneity between the two layers is a key factor for the appearance of
two peaks patterns.

All the numerical results are fully confirmed by an edge-based compartmental theory, see
our theoretical Eqs. (\ref{eq:19}) and (\ref{eq:20})
in {\sl Methods}. The solid curves in Fig. 3(a)-(d)
show the corresponding theoretical results.

These first numerics point that both a weak coupling and a difference in heterogeneity between the two layers
are necessary conditions for the emergence of the new pattern. An interesting question is whether the observed behavior
corresponds to a critical phenomenon. To figure out the answer, we let $\tau$ be the time interval between the two
peaks of $\rho_I$. In
particular, the two peaks will merge into a single one when $\tau=0$. Similarly, we let
$\delta_t=\mid t^B_{max}- t^A_{max}\mid$ be the time delay between the two peaks in layers $\mathcal{A}$
and $\mathcal{B}$, where $t^A_{max}$ and $t^B_{max}$ are the times of occurrence of the peak in $\rho_I^A$ and
$\rho_I^B$, respectively. The trivial situation would be that  for which $\tau=\delta_t$, but our numerical simulations
show that this condition is attained only
when $\tau$ is large, whereas one has $\tau<\delta_t$ when $\tau$ is sufficiently small. And, in particular, one has
$\delta_t>0$ when $\tau=0$.

Fig. 4(a) and (b) show the dependence of $\tau$ and $\delta_t$ on $\beta_{ab}$ for fixed
$\gamma=2.1$ and different $\langle k_{ab}\rangle$, respectively. From Fig. 4(a), one sees that when
$\langle k_{ab}\rangle$ is small, $\tau$ will decrease monotonically and be non-vanishing with the increasing of
$\beta_{ab}$, indicating that the event of two peaks always exists in the pattern. However, when $\langle k_{ab}\rangle$ is
increased, $\tau$ will decrease rapidly to zero, implying that there is a critical $\beta_{ab}^c$ for different
$\langle k_{ab}\rangle$. When $\beta_{ab}<\beta_{ab}^c$, the epidemic spreading event occurs through a pattern of two
peaks in the infected density, while for $\beta_{ab}>\beta_{ab}^c$ it occurs via the traditional
pattern with a single peak.
On its turn, Fig. 4(b) shows that $\delta_t$ decreases monotonically with the increase of $\beta_{ab}$ for all the
three cases of $\langle k_{ab}\rangle$, and it never vanishes. This is
because the spreading speed is different in homogeneous and heterogeneous networks. Generally speaking, epidemic
spreading is faster in heterogeneous network than in homogeneous network \cite{Barthelemy:2004}.

We then move to investigate the influence of the heterogeneity in degree distribution on the occurrence of the two
peaks pattern. Fig. 4(c) and (d) show
the dependence of $\tau$ and $\delta_t$ on $\beta_{ab}$ for different $\gamma$ and fixed $\langle k_{ab}\rangle=1.0$.
While the network heterogeneity decreases with the increasing of $\gamma$, it is easy to see that when $\gamma$ is large,
$\tau$ and $\delta_t$ decrease more prominently with $\beta_{ab}$.  Specifically, the
difference of spreading speeds between the two layers is not distinctive for large $\gamma$, resulting in the
disappearance of the two peaks. Therefore, increasing the coupling strength (i.e. $\langle k_{ab}\rangle$ and
$\beta_{ab}$) or decreasing the heterogeneity of network topology between the networks $\mathcal{A}$ and $\mathcal{B}$
will decrease the time delay of epidemic outbreak and then suppress the pattern of two peaks.

We have also confirmed all these numerical results in Fig. 4(a)-(d) by the theoretical Eqs. (\ref{eq:19}) and
(\ref{eq:20}) in {\sl Methods}. For each set of parameters in Fig. 4(a)-(d), we first produce the infected densities
$\rho_I, \rho_I^A$, and $\rho_I^B$ from Eqs. (\ref{eq:19}) and (\ref{eq:20}), as done in Fig. 3(a)-(d), and then
measure the corresponding $\tau$ and $\delta_t$. The solid curves in Fig. 4(a)-(d) show the theoretical $\tau$ and
$\delta_t$. One can easily see that the theoretical results are fully consistent with the numerical results.

\section*{Discussion}
Let us remark that the weak coupling condition  predicted by us for the occurrence of the novel epidemic
pattern is actually  consistent with the cases of the data of Fig. 1.
As it is well
known, indeed, Hong Kong in Fig. 1(a) consists of islands, and the movement of individuals between different islands
is not as convenient as that within each single island, and thus the coupling between neighboring islands can be
considered to be weak. At the same time, the population distribution in Hong Kong central island is significantly
different from that characterizing the surrounding islands, confirming the presence of the second ingredient predicted
by our theory: i.e. a difference in the heterogeneity of the layers' structures.

In Boston, a river separates the city
into two parts, which (to all extent) can be considered as equivalent to two islands. The same reasoning applies to
the neighboring cities of Bristol and Newcastle and the neighboring cities of Bristol and Sheffield. As Bristol and
Newcastle, Newcastle and Sheffield are separated regions in the United Kingdom, and they can therefore be considered
as a pair of weakly coupled networks.

Our predictions were obtained on coupled SF-RR networks, and it is legitimate to seek for generality of the two
peaks pattern phenomenon, by means of investigating coupled networks
with other topological structures. For this purpose, we have also studied the case of SF-SF and
RR-RR networks, respectively. Very interestingly, one finds that the pattern of two peaks can be still observed by
adjusting the coupling strength between the coupled layers (see Fig. 7 in SI for details).
On the other hand, extension to three-layered model was also considered, and it was found that
there is a small probability to produce a pattern of three peaks (see Fig. 8 in SI for details). Therefore, while
in principle one can expect a multi-peaks pattern to occur in a
multilayered network, the majority of unusual cases (i.e. cases in which the epidemic event is not happening with a
single maximum of infected density) will be characterized by just two peaks, in full consistency with the data of Fig. 1.

In summary, epidemic spreading has been well studied in the past decades but mainly focused on
outbreaks corresponding to patterns with a
single peak of infected density. We here reported (from real data) the evidence that also a pattern of two
peaks in a single epidemic period is possible.
We pointed out that such a pattern is a genuine product of a multi-layered interaction structure, and we have
introduced a proper model able to fully capture the mechanisms for its occurrence. Our model, together with reproducing 
the classical pattern of a single peak, can generate the pattern with two peaks when
proper conditions on weak coupling between the layers and difference in heterogeneity of the layers' structures are
satisfied.

\section*{Methods}

\subsection{A two-layered network epidemic model.}
We consider a two-layered network model with coupling between its two layers, i.e. the networks $\mathcal{A}$ and
$\mathcal{B}$ in Fig. 2. We let the two networks have the same size $N_a=N_b=N$ and their degree distributions
$P_A(k)$ and $P_B(k)$ be different. Each node has two kinds of links, i.e. intra-connection
(within $\mathcal{A}$ or $\mathcal{B}$) and interconnection between $\mathcal{A}$ and $\mathcal{B}$. The former
consists of the degree distributions of $P_A(k)$ and $P_B(k)$ while the latter gives rise to the interconnection network
$\mathcal{AB}$. We let $\langle k_a\rangle$, $\langle k_b\rangle$, and $\langle k_{ab}\rangle$ represent the average
degrees of networks $\mathcal{A}$, $\mathcal{B}$ and $\mathcal{AB}$, respectively. In details, we first generate
two separated networks $\mathcal{A}$ and $\mathcal{B}$ with the same size $N$ and different degree distributions
$P_A(k)$ and $P_B(k)$, respectively. Then, we add links between $\mathcal{A}$ and $\mathcal{B}$. That is, we randomly
choose two nodes from $\mathcal{A}$ and $\mathcal{B}$ and then connect them if they are not connected yet. The process
is repeated until all the needed specifications are attained.

In the above way, one obtains a uncorrelated two-layered network.

To discuss epidemic spreading in such a framework, we let each node represent a SIR model. In this model, a
susceptible node has two ways to be infected. One is from contacting with infected individuals in the network
$\mathcal{A}$ (or $\mathcal{B}$), represented by $\beta_a$ (or $\beta_b$). The other is from the coupled network 
$\mathcal{AB}$,
represented by $\beta_{ab}$ (see Fig. 2). Thus, a susceptible node will be infected with a probability $1-(1-\beta_a)^{k^{inf}}(1-\beta_{ab})^{k_{ab}^{inf}}$, where $k^{inf}$ is the infected neighbors in the same network
and $k_{ab}^{inf}$ is the infected neighbors in the coupled network. At the same time, an infected node will become
refractory by a probability $\mu$.

In our numerical simulations, we use scale-free (SF) and regular random (RR) graphs as networks $\mathcal{A}$
and $\mathcal{B}$, respectively. The network size is $N_a=N_b=10\,000$, the average degree
$\langle k_a \rangle=\langle k_b\rangle=6$, and initially $0.1\%$ of individuals of network $\mathcal{A}$
are chosen to be infected.

\subsection{Edge-based compartmental theory for a single network.}
Let us first illustrate the edge-based compartmental theory for a
single network, by following the methods and tools introduced in Refs.\cite{Volz:2008,Miller:2011,Shu:2016,Miller:2012,Volz:2011,Valdez:2012,Miller:2013a,Miller:2014,Miller:2013b,Wang:2016}.

For an uncorrelated, large and sparse network, the SIR model can be described in terms of the quantities $S(t)$, $I(t)$,
and $R(t)$, which represent the densities of the susceptible, infected, and recovered nodes at time $t$, respectively. Let
$\theta(t)$ be the probability that a neighbor $v$ of $u$ has not transmitted the disease to $u$ along the edge
connecting them up to time $t$. Then, the node $u$ with degree $k$ is susceptible at time $t$ as $s(k,t)=\theta(t)^k$.
Averaging over all $k$, the density of susceptible nodes at time $t$ is given by
\begin{equation}\label{eq:1}
S(t)=\sum_{k=0}^\infty P(k)\theta(t)^k
\end{equation}
where $P(k)$ is the degree distribution of the network. In order to solve for $S(t)$, one needs to know $\theta(t)$.
Since a neighbor $v$ of node $u$ may be susceptible, infected, or recovered, $\theta(t)$ can be expressed as
\begin{equation}\label{eq:2}
\theta(t)=\Phi^S(t)+\Phi^I(t)+\Phi^R(t)
\end{equation}
where $\Phi^S(t),\Phi^I(t),\Phi^R(t)$ is the probability that the neighbor $v$ is in the susceptible, infected,
recovery state, respectively,  and has not transmitted the disease to node $u$ through their connection. Once
these three parameters can be derived, we will get the density of susceptible nodes at time $t$
by substituting them into Eq. (\ref{eq:2}) and then into Eq. (\ref{eq:1}). To this
purpose, in the following, we will focus on how to solve them.

To find $\Phi^S(t)$, we now consider a randomly chosen node $u$, and
assume this node is in the cavity state, which means that it cannot transmit any disease to its neighbors $v$ but
can be infected by its neighbors. In this case, the neighbor $v$ can only get
the disease from its other neighbors except the node $u$. Thus,
node $v$ with degree $k'$ is susceptible with probability $\theta(t)^{k'-1}$ at time $t$ . For uncorrelated networks,
the probability that one edge from node $u$ connects with a node $v$ with degree $k'$ is
$k'P(k)/\langle k \rangle$. Summing over all possible $k'$, one obtains
\begin{equation}\label{eq:3}
\Phi^S(t)= \frac{\sum_{k'}k'P(k) \theta(t)^{k'-1}}{\langle k \rangle}
\end{equation}

According to the SIR spreading process, the growth of $\Phi^R(t)$ includes two consecutive events:
first, an infected neighbor has not transmitted the infection to
node $u$ via with probability $1-\beta$; second, the infected neighbor has been recovered with
probability $\mu$. Combining these two events, the $\Phi^I(t)$ to $\Phi^R(t)$ flux is $\mu(1-\beta)\Phi^I(t)$.
Thus, one gets
\begin{equation}\label{eq:4}
\frac{d\Phi^R(t)}{dt}= \mu(1-\beta)\Phi^I(t)
\end{equation}

Once the infected neighbor $v$ transmits the disease to $u$ successfully (with probability $\beta$),
the $\Phi^I(t)$ to $1-\theta(t)$ flux will be $\beta\Phi^I(t)$, which means
\begin{equation}\label{eq:5_1}
\frac{d(1-\theta(t))}{dt}=\beta\Phi^I(t) \nonumber
\end{equation}
That is
\begin{equation}\label{eq:5}
\frac{d\theta(t)}{dt}=-\beta\Phi^I(t)
\end{equation}
Combining Eqs. (\ref{eq:4}) and (\ref{eq:5}), and considering (as initial conditions) $\theta(0)=1$ and
$\Phi^R(0)=0$, one obtains
\begin{equation}\label{eq:6}
\Phi^R(t)=\frac{\mu(1-\theta(t))(1-\beta)}{\beta}
\end{equation}
Substituting Eqs. (\ref{eq:3}) and (\ref{eq:6}) into Eq. (\ref{eq:2}), one gets an expression for $\Phi^I(t)$ in
terms of $\theta(t)$, and then one can rewrite Eq. (\ref{eq:5}) as
\begin{equation}\label{eq:7}
\frac{d\theta(t)}{dt}=-\beta\theta(t)+\beta\frac{\sum_{k'}k'P(k) \theta(t)^{k'-1}}{\langle k \rangle}+\mu(1-\theta(t))(1-\beta)
\end{equation}

Thus, the equation of the system comes out to be
\begin{equation}\label{eq:8}
\frac{dR(t)}{dt}=\mu I(t), \quad  S(t)=\sum_{k=0}^\infty P(k)\theta(t)^k, \quad  I(t)=1-S(t)-R(t)
\end{equation}
In fact, Eq. (\ref{eq:7}) does not depend on Eq. (\ref{eq:8}), so the system is
governed by the single ordinary differential equation (\ref{eq:7}).
Although the resulting equation are simpler than those found by other methods,
it can be proven to exactly predict the disease dynamics in the large-population
limit for different network topologies\cite{Miller:2013a,Decreusefond:2012}.

\subsection{The theory for two-layered networks.}

When one assumes that the population is made up of two interacting networks, then $P_j(k_1,k_2)$ denote the probability
that a node of network $j$ has $k_1$ degree in network $1$ and $k_2$ in network $2$.
For the sake of simplicity, one can name the two
networks $\mathcal{A}$ and $\mathcal{B}$ as $1$ and $2$. Let $\beta_{j,l}$ be the rate of
transmission across an edge from network $l$ to network $j$, and let us define $\mu$ to be the recovery rate of a node
in any network.

$\theta_{j,l}$ can be defined to be the probability that an edge to a test node $u$ in network $j$ ($j=1,2$) is
coming from network $l$ ($l=1,2$), and has not transmitted the infection.

Now, $\theta_{1,2}$ can be solved as in the case of a single network. Since a neighbor $v$ in network 2 of node $u$ in
network 1 may be susceptible, infected, or recovered, $\theta_{1,2}$ can be expressed as
\begin{equation}\label{eq:9}
\theta_{1,2}=\Phi^S_{1,2}+\Phi^I_{1,2}+\Phi^R_{1,2}
\end{equation}
where $\Phi^S_{1,2}$, $\Phi^I_{1,2}$, $\Phi^R_{1,2}$ is the probability that the neighbor $v$ is in the
susceptible, infected, recovery state, and has not transmitted the disease to node $u$ through their connection.

Similarly, to find $\Phi^S_{1,2}$, the neighbor $v$ in network 1 can only get
the disease from its other neighbors except the node $u$ in network 2. Thus,
the node $v$ with degree $k_1$ in network 1 and degree $k_2$ in network 2 is susceptible with probability
$\theta_{2,1}^{k_1-1}\theta_{2,2}^{k_2}$ at time $t$. For uncorrelated networks, the probability that one
edge from node $u$ connects with a node $v$ with degree $(k_1,k_2)$ is
$\frac{k_1P_2(k_1,k_2)}{\sum_{k_1,k_2}k_1P_2(k_1,k_2)}$. Thus, one has
\begin{eqnarray}\label{eq:10}
\Phi^S_{1,2}&=& \frac{\sum_{k_1,k_2}k_1P_2(k_1,k_2) \theta_{2,1}^{k_1-1}\theta_{2,2}^{k_2}}{\sum_{k_1,k_2}k_1P_2(k_1,k_2)}
\end{eqnarray}

It is easily to know that the growth of $\Phi^R_{1,2}$ includes two consecutive events:
first, an infected neighbor has not transmitted the infection to
node $u$ via with probability $1-\theta_{1,2}$; second, the infected neighbor has been recovered with
probability $\mu$. Combining these two events, the $\Phi^I_{1,2}$ to $\Phi^R_{1,2}$ flux is $\mu(1-\theta_{1,2})\Phi^I_{1,2}$.
Thus, one gets
\begin{eqnarray}\label{eq:11}
\frac{d\Phi^R_{1,2}}{dt}&=& \mu(1-\theta_{1,2})\Phi^I_{1,2}
\end{eqnarray}

Once the infected neighbor $v$ in network 1 transmits the disease to node $u$ in network 2 successfully (with
probability $\beta_{1,2}$), the $\Phi^I_{1,2}$ to $1-\theta_{1,2}$ flux will be $\beta_{1,2}\Phi^I_{1,2}$, which means
\begin{eqnarray}\label{eq:12}
\frac{d\theta_{1,2}}{dt}&=&-\beta_{1,2}\Phi^I_{1,2}
\end{eqnarray}

Combining Eqs. (\ref{eq:11}) and (\ref{eq:12}), and considering (as initial conditions) $\theta_{1,2}(0)=1$ and
$\Phi^R_{1,2}(0)=0$, one obtains
\begin{eqnarray}\label{eq:13}
\Phi^R_{1,2}&=&\frac{\mu(1-\theta_{1,2})(1-\beta_{1,2})}{\beta_{1,2}}
\end{eqnarray}
So, one gets
\begin{eqnarray}\label{eq:15}
\dot{\theta}_{1,2}&=&-\beta_{1,2}(\theta_{1,2}-\Phi^S_{1,2}-\Phi^R_{1,2})  \nonumber\\
 &=&-\beta_{1,2}\theta_{1,2}+\beta_{1,2} \frac{\sum_{k_1,k_2}k_1P_2(k_1,k_2) \theta_{2,1}^{k_1-1}\theta_{2,2}^{k_2}}{\sum_{k_1,k_2}k_1P_2(k_1,k_2)}+\mu(1-\theta_{1,2})(1-\beta_{1,2})
\end{eqnarray}

Similarly, one can write down $\theta_{1,1}$, $\theta_{2,1}$ and $\theta_{2,2}$ as follows
\begin{eqnarray}
\dot{\theta}_{1,1}=-\beta_{1,1}\theta_{1,1}+\beta_{1,1} \frac{\sum_{k_1,k_2}k_1P_1(k_1,k_2) \theta_{1,1}^{k_1-1}\theta_{1,2}^{k_2}}{\sum_{k_1,k_2}k_1P_1(k_1,k_2)}+\mu(1-\theta_{1,1})(1-\beta_{1,1}) \label{eq:16} \\
\dot{\theta}_{2,1} = -\beta_{2,1}\theta_{2,1}+\beta_{2,1} \frac{\sum_{k_1,k_2}k_2P_1(k_1,k_2) \theta_{1,1}^{k_1}\theta_{1,2}^{k_2-1}}{\sum_{k_1,k_2}k_2P_1(k_1,k_2)}+\mu(1-\theta_{2,1})(1-\beta_{2,1}) \label{eq:17}\\
\dot{\theta}_{2,2}=-\beta_{2,2}\theta_{2,2}+\beta_{2,2} \frac{\sum_{k_1,k_2}k_2P_2(k_1,k_2) \theta_{2,1}^{k_1}\theta_{2,2}^{k_2-1}}{\sum_{k_1,k_2}k_2P_2(k_1,k_2)}+\mu(1-\theta_{2,2})(1-\beta_{2,2}) \label{eq:18}
\end{eqnarray}
With Eqs. (\ref{eq:15}-\ref{eq:18}) on hand, the densities associated with each distinct state can be obtained by
\begin{eqnarray}
\dot{R}_1=\mu I_1(t), \quad  S_1(t)=\sum_{k_1,k_2}^\infty P_1(k_1,k_2)\theta_{1,1}^{k_1}\theta_{1,2}^{k_2},  \quad  I_1(t)=1-S_1(t)-R_1(t) \label{eq:19}\\
\dot{R}_2=\mu I_2(t), \quad  S_2(t)=\sum_{k_1,k_2}^\infty P_2(k_1,k_2)\theta_{2,2}^{k_2}\theta_{2,2}^{k_1},  \quad  I_2(t)=1-S_2(t)-R_2(t) \label{eq:20}
\end{eqnarray}

Eqs. (\ref{eq:19}) and (\ref{eq:20}) are the main theoretical results from which the theoretical curves in Figs. 3
and 4 are calculated. Furthermore, we find that the threshold for the whole network to show epidemic outbreak can be
theoretically figured out by the Jacobian matrix $\textbf{J}$ of Eqs. (\ref{eq:15}-\ref{eq:18}) (see Figs. 5 and 6 in
SI for details). Especially, when the coupling between the two layers is very weak, the obtained threshold will be
consistent with the previous findings \cite{Mendiola:2012,Sahneh:2013}.

\begin{addendum}

\item[Acknowledgments]
Authors acknowledge the Centre for Health Protection, Department of Health, the Government
of the Hong Kong Special Administrative Region, and the USA National
Notifiable Diseases Surveillance System as digitized by Project Tycho for
providing data. This work was partially supported by the NNSF of China under Grant
Nos. 11135001, 11375066, 973 Program under Grant No.
2013CB834100.

\item[Author Contributions]
M.Z. and Z.L. conceived the research project.
M.Z., W. W., M. T. J. Z. and Z.L. performed the research. All Authors analyzed the results. M.Z., S. B. and Z.L.
wrote the paper. All Authors reviewed the Manuscript.

\item[Competing Interests]
Authors declare no competing financial interests.

\item[Correspondence] Correspondence and requests for materials
should be addressed to Z.L.~(zhliu@phy.ecnu.edu.cn).

\end{addendum}

\newpage
\begin{figure}
\centering
\includegraphics[width=\linewidth]{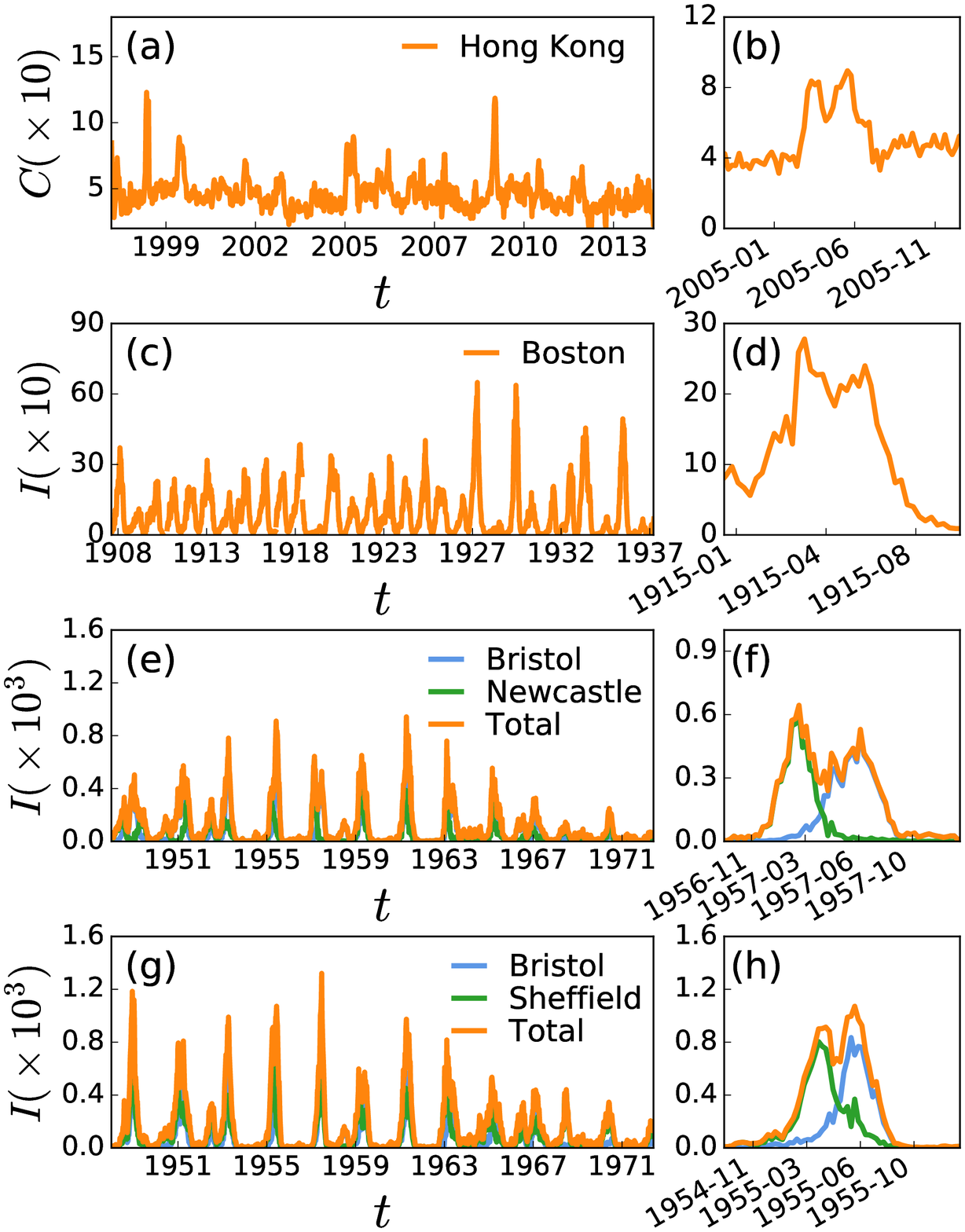}
\caption{(Color online). Time series of recurrent epidemics in different cities over the world.
(a) The weekly consultation rates of influenza-like illness (per 1,000 consultations) from $1998/1/3$
to $2014/8/2$ in Hong Kong for the General Practitioners (GP) sentinel system. (b) Zoom of one of the patterns
with two peaks, occurring at around $2005/6$ in (a).
(c) The time series of reported weekly measles infective cases I in Boston. (d) Zoom of one of the patterns
with two peaks, occurring at around $1915/4$ in (c).
(e)-(h): Time series of infectious disease in two coupled cities.
(e) and (f): The yellow line represents the total number of weekly measles infective cases in coupled Bristol
and Newcastle, while the blue and green lines represent the number in Bristol and Newcastle, respectively. (f)
is one of the patterns, occurring at around $1957/3$ in (e).
(g) and (h): The yellow line represents the total number of weekly measles infective cases in coupled Bristol
and Sheffield, while the blue and green lines represent the number in Bristol and Sheffield, respectively. (h)
is one of the patterns, occurring at around $1955/5$ in (g).
}
\label{Fig:data}
\end{figure}

\begin{figure}
\centering
\includegraphics[width=\linewidth]{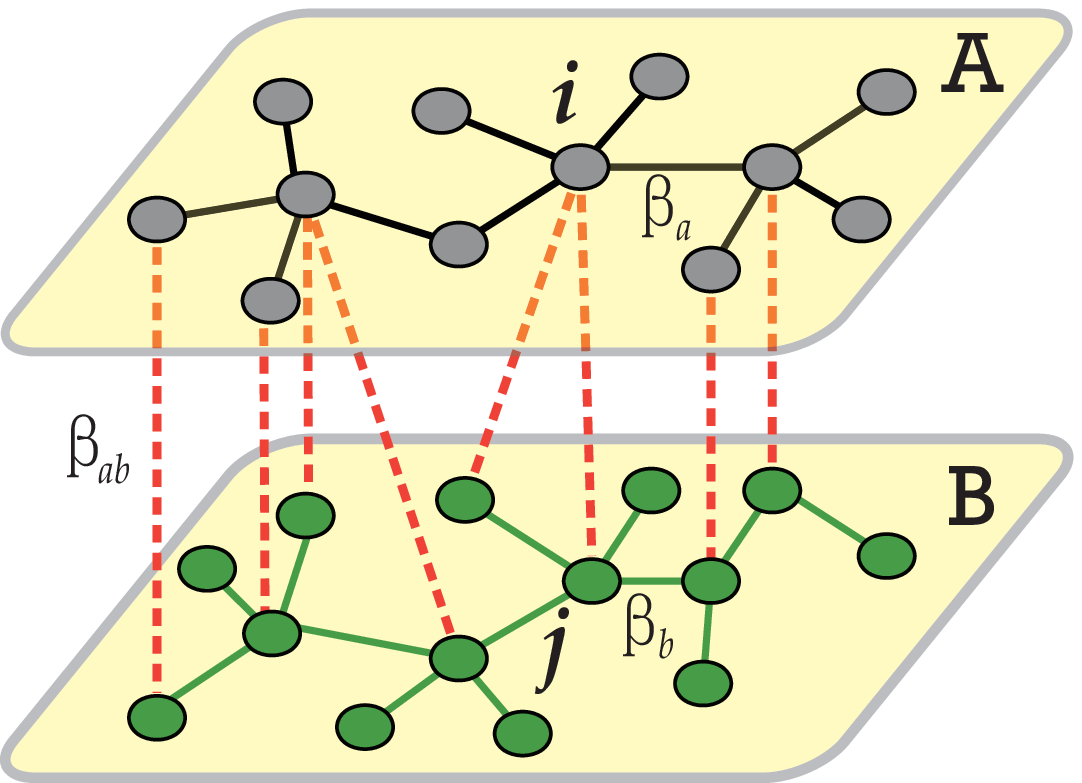}
\caption{(Color online). Sketch of the two-layered network model, which reproduces the pattern of two
peaks. ``Black", ``green" and ``red" lines represent the links of the networks $\mathcal{A}$,
$\mathcal{B}$ and the inter-network $\mathcal{AB}$, respectively. $\beta_a$, $\beta_b$ and $\beta_{ab}$
denotes the infectious rates of networks $\mathcal{A}$, $\mathcal{B}$ and $\mathcal{AB}$.  }
\label{Fig:model}
\end{figure}

\begin{figure}
\centering
\includegraphics[width=\linewidth]{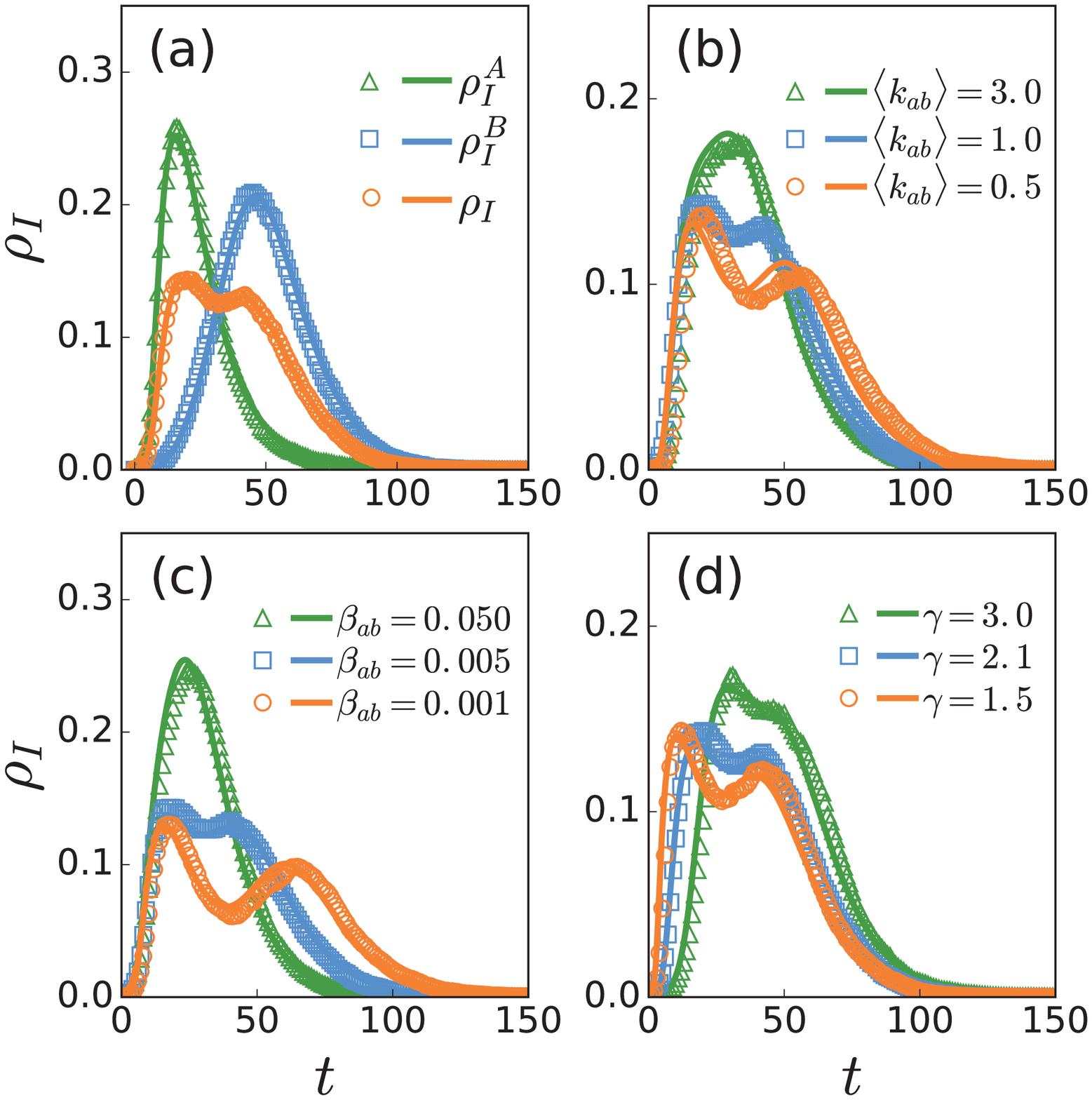}
\caption{(Color online). The two peaks pattern occurring in different conditions, with $\mu=0.1$,
$\beta_a=\beta_b=0.05$, $\langle k_a\rangle=\langle k_b\rangle=6$, and $N_a=N_b=10,000$, where the symbols
represent the simulated results and the lines denote the corresponding theoretical predictions (calculated via
the edge-based compartmental theory described in the Method section).
(a) $\rho_I(t)$ vs. $t$, where $\rho_I$, $\rho_I^A$ and $\rho_I^B$ represent the infected densities in the
entire network, and the networks $\mathcal{A}$ and $\mathcal{B}$, respectively. Other parameters are
$\gamma=2.1$, $\langle k_{ab}\rangle=1.0$, and $\beta_{ab}=0.005$.
(b) The influence of $\langle k_{ab}\rangle$ on $\rho_I(t)$ with $\gamma=2.1$ and $\beta_{ab}=0.005$, where the
``triangles", ``squares" and ``circles" represent the cases of $\langle k_{ab}\rangle=3.0, 1.0$ and $0.5$,
respectively.
(c) The influence of $\beta_{ab}$ on $\rho_I(t)$ with $\gamma=2.1$ and $\langle k_{ab}\rangle=1.0$, where the
``triangles", ``squares" and ``circles" represent the cases of $\beta_{ab}=0.05, 0.005$ and $0.001$, respectively.
(d) The influence of $\gamma$ on $\rho_I(t)$ with $\beta_{ab}=0.005$ and $\langle k_{ab}\rangle=1.0$, where the
``triangles", ``squares" and ``circles" represent the cases of $\gamma=3.0, 2.1$ and $1.5$, respectively.
}
\label{Fig:model_timeseeries}
\end{figure}

\begin{figure}
\centering
\includegraphics[width=\linewidth]{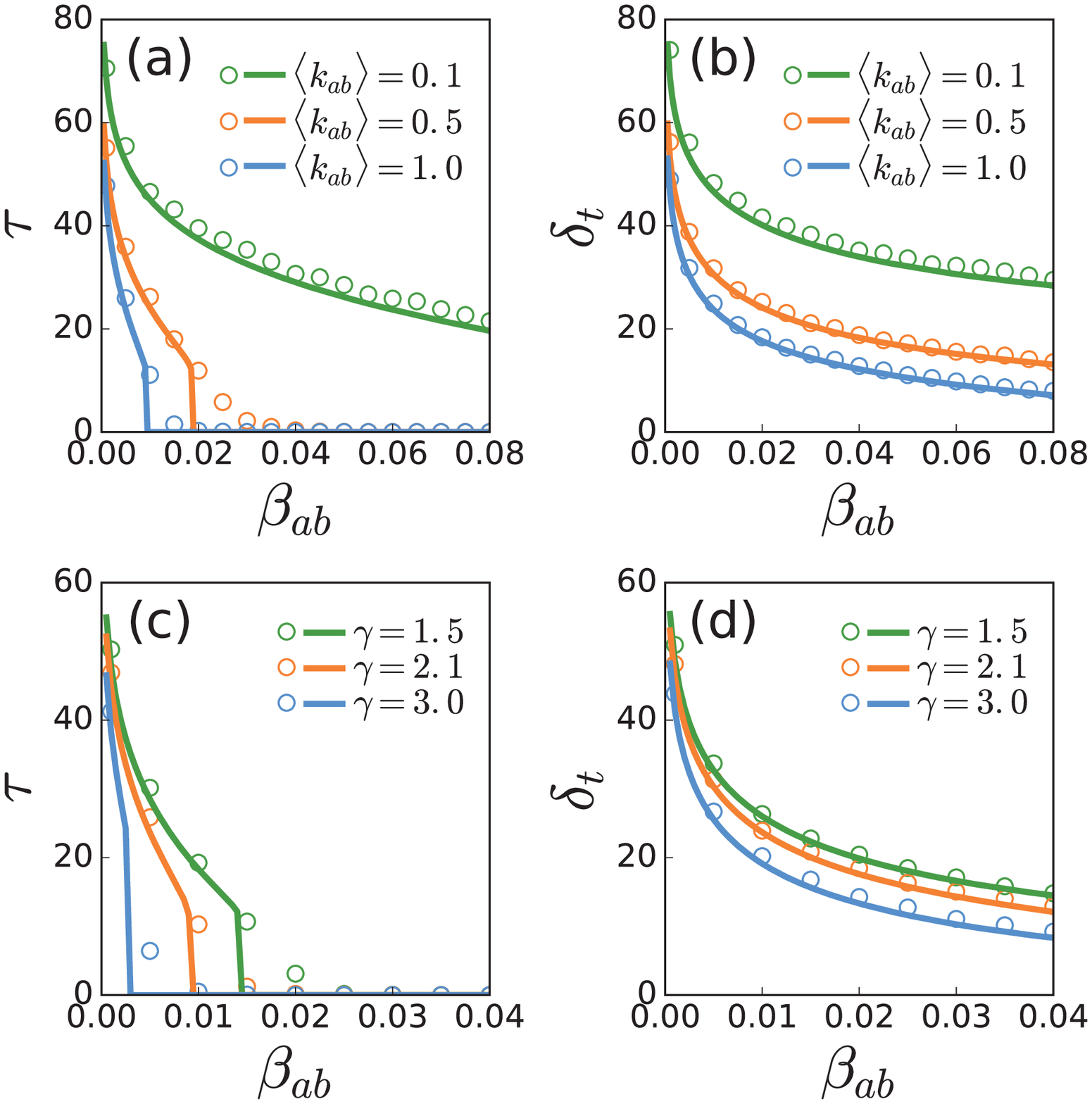}
\caption{(Color online).
(a) and (b) report the dependence of $\tau$ and $\delta_t$ on $\beta_{ab}$ (see main text for definitions) with different
$\langle k_{ab}\rangle$ in SF-RR networks.
(c) and (b) report the dependence of $\tau$ and $\delta_t$ on $\beta_{ab}$ with different
$\gamma$ in SF-RR networks. Symbols represent the simulated results and the lines are the
corresponding theoretical results (calculated via
the edge-based compartmental theory described in the Method section).
Other parameters as in the caption of Fig. 3. All the results are averaged over $100$ independent realizations.
}
\label{Fig:phase}
\end{figure}


\section*{References}

\begin{thebibliography}{}

\bibitem{Pastor:2015}
Pastor-Satorras R., Castellano C., Van Mieghem P., \& Vespignani A.
Epidemic processes in complex networks.
\emph{Rev. Mod. Phys.}, \textbf{87}(3), 925 (2015).

\bibitem{Barrat:2008}
Barrat A., Barthelemy M., \& Vespignani A.
\emph{Dynamical processes on complex networks. }
(Cambridge University Press, Cambridge, England, 2008).

\bibitem{Wang:2017}
Wang W., Tang M., Stanley H. E., \& Braunstein L. A.
Unification of theoretical approaches for epidemic spreading on complex networks.
\emph{Rep. Prog. Phys.}, \textbf{80}(3), 036603 (2017).

\bibitem{Salehi:2015}
Salehi M., Sharma R., Marzolla M., Magnani M., Siyari P., \& Montesi D.
Spreading processes in multilayer networks.
\emph{IEEE Transactions on Network Science and Engineering}, \textbf{2}(2), 65-83 (2015).

\bibitem{Pastor-Satorras:2001}
Pastor-Satorras R., \& Vespignani A.
Epidemic spreading in scale-free networks.
\emph{Phys. Rev. Lett.}, {\bf 86}, 3200 (2001).

\bibitem{Boguna:2002}
Bogu\~{n}\'{a} M., \& Pastor-Satorras R.
Epidemic spreading in correlated complex networks.
\emph{Phys. Rev. E}, {\bf 66}, 047104 (2002).

\bibitem{Colizza:2007a}
Colizza V., Pastor-Satorras R. \& Vespignani A.
Reaction-diffusion processes and metapopulationmodels in heterogeneous networks.
{\it Nat. Phys.}, {\bf 3}, 276-282 (2007).

\bibitem{Colizza:2007b}
Colizza V. \& Vespignani A. Invasion Threshold in Heterogeneous Metapopulation Networks.
\emph{Phys. Rev. Lett.}, {\bf 99}, 148701 (2007).

\bibitem{Andrea:2008}
Baronchelli A., Catanzaro M. \& Pastor-Satorras R. Bosonic reaction-diffusion processes on scale-free networks.
\emph{Phys. Rev. E}, {\bf 78}, 016111 (2008).

\bibitem{Liu:2009}
Tang M., Liu L. \& Liu Z. Influence of dynamical condensation on epidemic spreading in scale-free networks.
\emph{Phys. Rev. E}, {\bf 79}, 016108 (2009).

\bibitem{Vazquez:2007}
Vazquez A., Racz B., Lukacs A., \& Barabasi A. L.
Impact of non-Poissonian activity patterns on spreading processes.
\emph{Phys. Rev. Lett.}, {\bf 98}, 158702 (2007).

\bibitem{Meloni:2009}
Meloni S., Arenas A., \& Moreno Y.
Traffic-driven epidemic spreading in finite-size scale-free networks.
\emph{Proc. Natl. Acad. Sci. USA }, {\bf 106}, 16897-16902 (2009).

\bibitem{Balcan:2009}
Balcan D., Colizza V., Gon\c{c}alves B., Hu H., Ramasco J. J., \& Vespignani A.
Multiscale mobility networks and the spatial spreading of infectious diseases.
\emph{Proc. Natl. Acad. Sci. USA }, {\bf 106}, 21484-21489 (2009).

\bibitem{Ruan:2012}
Ruan Z., Tang M., \& Liu Z.
Epidemic spreading with information-driven vaccination.
\emph{Phys. Rev. E}, {\bf 86}, 036117 (2012).

\bibitem{Liu:2012}
Liu S., Perra N., Karsai M., \& Vespignani A.
Controlling contagion processes in activity driven networks.
\emph{Phys. Rev. Lett.}, {\bf 112}, 118702 (2014).

\bibitem{Tang:2009}
Tang M., Liu Z., \& Li B.
Epidemic spreading by objective traveling.
\emph{Europhys. Lett.}, {\bf 87}, 18005 (2009).

\bibitem{Liu:2010}
Liu Z.
Effect of mobility in partially occupied complex networks.
\emph{Phys. Rev. E}, {\bf 81}, 016110 (2010).

\bibitem{Gross:2006}
Gross T., D¡¯Lima C. J. D., \& Blasius B.,
Epidemic dynamics on an adaptive network.
\emph{Phys. Rev. Lett.}, \textbf{96}(20), 208701 (2006).

\bibitem{Gross:2008}
Gross T., \& Kevrekidis I. G.
Robust oscillations in SIS epidemics on adaptive networks: Coarse graining by automated moment closure.
\emph{Europhys. Lett.}, \textbf{82}(3), 38004 (2008).

\bibitem{Zhou:2012}
Zhou J., Xiao G., Cheong S. A., Fu X., Wong L., Ma S., \& Cheng T. H.
Epidemic reemergence in adaptive complex networks.
\emph{Phys. Rev. E}, \textbf{85}(3), 036107 (2012).

\bibitem{Zhou:2013}
Zhou J., Xiao G., \& Chen G.
Link-based formalism for time evolution of adaptive networks.
\emph{Phys. Rev. E}, \textbf{ 88}(3), 032808 (2013).

\bibitem{Marceau:2010}
Marceau V., No\"{e}l P. A., H\'{e}bert-Dufresne L., Allard A., \& Dub\'{e} L. J.
Adaptive networks: Coevolution of disease and topology.
\emph{Phys. Rev. E}, \textbf{82}(3), 036116 (2010).

\bibitem{Boccaletti:2014}
Boccaletti S., Bianconi G., Criado R., del Genio C. I., G\'{o}mez-Garde\~{n}es J., Romance M.,
Sendi\~{n}a-Nadal I., Wang Z., \& Zanin M. The structure and dynamics of multilayer networks.
\emph{Phys. Rep.} \textbf{544}, 1 (2014).


\bibitem{Feng:2015}
Feng L., Monterola C. P., \& Hu Y.
The simplified self-consistent probabilities method for percolation and its application to interdependent networks.
\emph{New J. Phys.}, \textbf{17}(6), 063025 (2015).

\bibitem{Sahneh:2013}
Sahneh F. D., Scoglio C., \& Chowdhury F. N.
Effect of coupling on the epidemic threshold in interconnected complex networks: A spectral analysis.
\emph{In 2013 American Control Conference} (pp. 2307-2312). IEEE (2013).

\bibitem{Wang:2013}
Wang H., Li Q., D'Agostino G., Havlin S., Stanley H. E., \& Van Mieghem P.
Effect of the interconnected network structure on the epidemic threshold.
\emph{Phys. Rev. E}, \textbf{88}(2), 022801 (2013).

\bibitem{Yagan:2013}
Yagan O., Qian D., Zhang J., \& Cochran D.
Conjoining speeds up information diffusion in overlaying social-physical networks.
\emph{IEEE J. Sel. Areas Commun.}, \textbf{31}(6), 1038-1048 (2013).

\bibitem{Newman:2005}
Newman M. E.
Threshold effects for two pathogens spreading on a network.
\emph{Phys. Rev. Lett.}, \textbf{95}(10), 108701 (2005).
\bibitem{Marceau:2011}
Marceau V., No\"{e}l P. A., H\'{e}bert-Dufresne L., Allard A., \& Dub\'{e} L. J.
Modeling the dynamical interaction between epidemics on overlay networks.
\emph{Phys. Rev. E}, \textbf{84}, 026105 (2011).

\bibitem{Buono:2015}
Buono C., \& Braunstein L. A.
Immunization strategy for epidemic spreading on multilayer networks.
\emph{Europhy. Lett.}, \textbf{109}(2), 26001 (2015).

\bibitem{Buono:2014}
Buono C., Alvarez-Zuzek L. G., Macri P. A., \& Braunstein L. A.
Epidemics in partially overlapped multiplex networks.
\emph{PloS One}, \textbf{9}(3), e92200 (2014).

\bibitem{Zhao:2014}
Zhao Y., Zheng M., \& Liu Z.
A unified framework of mutual influence between two pathogens in multiplex networks.
\emph{Chaos}, {\bf 24}, 043129 (2014).

\bibitem{Holme:2012}
Holme P., \& Saram{\"a}ki J. Temporal networks.
{\it Phys. Rep.}, {\bf 519}, 97-125 (2012).

\bibitem{Perra:2012}
Perra N., Goncalves B., Pastor-Satorras R. \& Vespignani A. Activity driven modeling of time varying networks.
{\it Sci. Rep.} {\bf 2}, 469 (2012).

\bibitem{www:hongkong}
Department of Health, Hong Kong. Weekly consultation rates of influenza-like illness data.
http://www.chp.gov.hk/en/sentinel/26/44/292.html. Date of access: 15/06/2014.

\bibitem{Zheng:2015}
Zheng M., Wang C., Zhou J., Zhao M., Guan S., Zou Y., \& Liu Z.
Non-periodic outbreaks of recurrent epidemics and its network modelling.
\emph{Sci. Rep.} \textbf{ 5}, 16010 (2015).

\bibitem{measles:2016}
The USA National Notifiable Diseases Surveillance System. Weekly measles infective cases.
http://www.tycho.pitt.edu/. Date of access: 04/08/2016.

\bibitem{Scarpino:2016}
Scarpino S. V., Allard A., \& H\'{e}bert-Dufresne L.
The effect of a prudent adaptive behaviour on disease transmission.
\emph{Nat. Phys.} \textbf{3832}, 1745-2481 (2016).

\bibitem{www:measdata}
http://ms.mcmaster.ca/~bolker/measdata.html

\bibitem{Catanzaro:2005}
Catanzaro M., Boguna M., \& Pastor-Satorras R. Generation of uncorrelated random scale-free networks.
{\it Phys. Rev. E} {\bf 71}, 027103 (2005).

\bibitem{Rucinski:2012}
Ruci\'{n}ski A., \& Wormald N. C. Random graph processes with degree restrictions.
\emph{Combinatorics, Probability and Computing}, \textbf{1}(2), 169-180 (1992).




\bibitem{Barthelemy:2004}
Barth\'{e}lemy M., Barrat A., Pastor-Satorras R., \& Vespignani A.
Velocity and hierarchical spread of epidemic outbreaks in scale-free networks.
\emph{Phys. Rev. Lett.}, \textbf{92}(17), 178701 (2004).

\bibitem{Volz:2008}
Volz E.
SIR dynamics in random networks with heterogeneous connectivity.
\emph{J. Math. Biol.}, \textbf{56}(3), 293-310 (2008).

\bibitem{Miller:2011}
Miller J. C.
A note on a paper by Erik Volz: SIR dynamics in random networks.
\emph{J. Math. Biol.}, \textbf{62}(3), 349-358 (2011).


\bibitem{Shu:2016}
Shu P., Wang W., Tang M., Zhao P., \& Zhang Y. C.
Recovery rate affects the effective epidemic threshold with synchronous updating.
\emph{Chaos}, \textbf{26}(6), 063108 (2016).

\bibitem{Miller:2012}
Miller J. C., Slim A. C., \& Volz E. M.
Edge-based compartmental modelling for infectious disease spread.
\emph{J R Soc Interface}, \textbf{9}(70), 890-906 (2012).

\bibitem{Volz:2011}
Volz E. M., Miller J. C., Galvani A., \& Meyers L. A.
Effects of heterogeneous and clustered contact patterns on infectious disease dynamics.
\emph{PLoS Comput Biol.}, \textbf{7}(6), e1002042 (2011).

\bibitem{Valdez:2012}
Valdez L. D., Macri P. A., \& Braunstein L. A.
Temporal percolation of the susceptible network in an epidemic spreading.
\emph{PLoS One}, \textbf{7}(9), e44188 (2012).

\bibitem{Miller:2013a}
Miller J. C., \& Volz E. M.
Incorporating disease and population structure into models of SIR disease in contact networks.
\emph{PloS One}, \textbf{8}(8), e69162 (2013).

\bibitem{Miller:2014}
Miller J. C.
Epidemics on networks with large initial conditions or changing structure.
\emph{PloS one}, \textbf{9}(7), e101421 (2014).

\bibitem{Miller:2013b}
Miller J. C.
Cocirculation of infectious diseases on networks.
\emph{Phys. Rev. E}, \textbf{87}(6), 060801 (2013).

\bibitem{Wang:2016}
Wang W., Tang M., Shu P., \& Wang, Z.
Dynamics of social contagions with heterogeneous adoption thresholds: Crossover phenomena in phase transition.
\emph{New J. Phys.}, \textbf{18}(1), 013029 (2016).

\bibitem{Decreusefond:2012}
Decreusefond L., Dhersin J. S., Moyal P., \& Tran V. C.
Large graph limit for an SIR process in random network with heterogeneous connectivity.
\emph{ Ann. Appl. Probab.}, \textbf{22}(2), 541-575 (2012).

\bibitem{Mendiola:2012}
Saumell-Mendiola A., Serrano M. A., \& Bogun\'{a} M.
Epidemic spreading on interconnected networks.
\emph{Phys. Rev. E} \textbf{86}, 026106 (2012).




\end{thebibliography}
\end{document}